----------------------------- Start of body part 2

\documentstyle[12pt]{article}

\begin{document}

\begin{titlepage}
\title{On the integrability of N=2 Landau-Ginzburg
models: A graph generalization of the Yang-Baxter equation}

\author{C\'esar G\'omez$^{1*}$ and Germ\'an Sierra$^2$\thanks{Permanent
address: Instituto de Matem\'aticas y F\'{\i}sica
Fundamental, CSIC; Serrano 123, E--28006 Spain.} \\
$^1$ {\it D\'epartement de Physique Th\'eorique, Universit\'e de Gen\`eve} \\
{\it CH--1211 Gen\`eve 4, Switzerland} \\
e--mail: GOMEZC@SC2A.UNIGE.CH \\
\and
$^2$ {\it Theory Division, CERN} \\
{\it CH--1211 Gen\'eve 23, Switzerland} \\
e--mail: SIERRA@CC.CSIC.ES}

\date{}

\maketitle

\begin{abstract}
The study of the integrability properties of the N=2 Landau-Ginzburg
models leads naturally to a graph generalization of the Yang-Baxter
equation which synthetizes the well known vertex and RSOS Yang-Baxter
equations.A non trivial solution of this equation is found for
the $t_2$ perturbation of the A-models, which turns out to be intimately
related to the Boltzmann weights of a Chiral-Potts model.
\end{abstract}

\vskip-18.0cm
\rightline{{\bf CERN-TH.6963/93}}
\rightline{{\bf UGVA 07/622/93}}
\rightline{{\bf August 1993}}
\vskip2cm

\end{titlepage}

\section*{Introduction}

N=2 Landau-Ginzburg models have been extensively used to describe
relevant and SUSY preserving perturbations of N=2 superconformal
field theories \cite{Va}. The resulting massive field theories
are in some cases integrable and the associated
scattering S-matrices satisfy the bootstrap and factorization
equations \cite{Za,Fe} . Those Landau-Ginzburg superpotentials which
are Morse functions contain in their spectrum all the
soliton configurations which interpolate between the different
critical points. Evenmore in some special cases a closed
scattering theory ,in the sense of bootstrap \cite{ZZ}, can be defined
choosing as fundamental particles a set of Bogomolnyi
solitons. Two main questions arise in the study of
N=2 Landau-Ginzburg models: a) when the massive theory
defined by a given non degenerated superpotential is
integrable and b) which is the associated scattering
S-matrix which satisfy the requirements of bootstrap
and factorization.

During the last few years these two questions have been
considered, which much or less succes, under different points
of view: Toda theories \cite{To}, quantum affine algebras \cite{Qa},
holomorphic field theories \cite{Ho}, graph rings \cite{Ga}.
In this letter we shall present an algebraic framework
which we hope will capture the integrability of the
N=2 Landau-Ginzburg models. The cornerstone of our
approach is a graph generalization of the Yang-Baxter
equation, whose non triviality is illustrated with
the $t_2$ perturbation of the $A$-models.

\section*{N=2 Landau-Ginzburg models: Graph Quantum Groups}

The Landau-Ginzburg models that we shall consider have
a N=2 supersymmetry algebra generated by the SUSY
charges $ Q^{\pm} , \bar{Q}^{\pm} $ together with the
topological charges $T , \bar{T}$,
fermion number operator $ \cal{F}$ and momentum operators
$P , \bar{P}$ satisfying the relations:

\begin{eqnarray}
( Q^{\pm})^2=( \bar{Q}^{\pm})^2= &
\{Q^+, \bar{Q}^- \}=  \{Q^-,
 \bar{Q}^+ \}= & 0 \nonumber \\
\{ Q^+,{Q}^- \}= P, & \{ \bar{Q}^+,
 \bar{Q}^- \}= \bar{P} &
\label{ 1 } \\
\{ Q^+, \bar{Q}^+ \} = T,  &
\{ {Q}^-, \bar{Q}^- \}= \bar{T} & \nonumber \\
\left[ {\cal F},Q^{\pm} \right]=\pm Q^{\pm}, &
\left[ {\cal F},\bar{Q}^{\pm} \right]=\mp \bar{Q}^{\pm}
& \nonumber
\end{eqnarray}

\noindent
Each LG model is characterized by a superpotential whose
critical points define the vertex of a graph. The links
of this graph are associated to the fundamental Bogomolnyi
solitons\footnote{ The graph we are refering to is
the same used in \cite{Ho} for the classification
of N=2 theories and it is related in some cases
with the dual
of the normalized ring in the sense of reference \cite{Ga}}.

Let us denote by $a, b, \cdots$ the vertices of the LG-graph,
then a Bogomolnyi soliton associated to the link $ (a,b)$
and rapidity $\theta$
give rise to a two dimensional irrep $\pi_{a,b}(\theta)$
of the N=2 algebra ( \ref{ 1 } ) given by:

\begin{eqnarray}
\pi_{a,b}(\theta)(Q^-) = \left( \begin{array}{ccc}
0 & 0 \\ \sqrt{m_{a,b}} e^{\theta/2}  & 0 \end{array} \right),
& &
\pi_{a,b}(\theta)(Q^+)= \left( \begin{array}{ccc}
0 &  \sqrt{m_{a,b}} e^{\theta/2} \\ 0 & 0  \end{array} \right)
\nonumber \\
 & & \label{ 2 } \\
\pi_{a,b}(\theta)(\bar{Q}^+) = \left( \begin{array}{ccc}
0 & 0 \\ \omega_{a,b}  \sqrt{m_{a,b}} e^{-\theta/2}  & 0 \end{array} \right),
& &
\pi_{a,b}(\theta)(\bar{Q}^-)= \left( \begin{array}{ccc}
0 & \omega_{a,b}^*  \sqrt{m_{a,b}} e^{-\theta/2} \\ 0 & 0  \end{array} \right)
\nonumber
\end{eqnarray}

\noindent
where

\begin{eqnarray}
m_{a,b} = 2 | \Delta_{a,b} |, & \omega_{a,b} = \frac{
\Delta_{a,b} }{ | \Delta_{a,b} | }, & \Delta_{a,b}= W_a -W_b
\label{3}
\end{eqnarray}

\noindent
$ W_a $ and $ W_b$ in these equations denotes the values of the
superpotential $W$ at the critical points $a$ and $b$ respectively.
The fermion number
$f_{a,b}$ of the soliton connecting these critical points
is obtained through the Hessian $H(x)$ of the
superpotential  $W(x)$ as:

\begin{equation}
exp( 2 \pi {\rm i} f_{a,b}) = phase \left( \frac{ det H(b)}{det H(a)}
\right)
\label{3b}
\end{equation}

Notice that these formula gives the soliton fermion number only modulo 1.

The N=2 algebra ( \ref{ 1 }) admits a bialgebra structure
given by the following comultiplication rules:

\begin{eqnarray}
\Delta Q^{\pm} & = & Q^{\pm} \otimes {\bf 1} +
e^{ \pm {\rm i} \pi \cal{F} } \otimes Q^{\pm} \label{4} \\
\Delta \bar{Q}^{\pm} & = & \bar{Q}^{\pm} \otimes {\bf 1} +
e^{ \mp {\rm i} \pi \cal{F} } \otimes \bar{Q}^{\pm} \nonumber
\end{eqnarray}

{}From the theory of quantum groups we know that non trivial
comultiplication rules, as the one shown above, yield
non trivial intertwiners. The intertwiner for the
N=2 LG models turn out to
have a rather rich structure, made manifest by the proliferation
of indices. Indeed a generic N=2 LG intertwiner depends on two sort
of indices: the LG indices $ a,b, \cdots $ which label the
vacua of the LG superpotential and the N=2 indices $i,j, \cdots$
which may take two values, say 0 or 1, depending on the state
in the N=2 supermultiplet of the Bogomolnyi soliton. Hence the
structure of the intertwiners of any N=2 model is condensed in the
expresion:

\begin{equation}
S^{k \; \ell}_{i \; j} \left(
\begin{array}{cc} a  & d   \\ b   & c   \end{array}
\right) : {\cal V}_{a,b}( \theta_1 ) \otimes
{\cal V}_{b,c}(\theta_2) \rightarrow
{\cal V}_{a,d}( \theta_2 ) \otimes
{\cal V}_{d,c}(\theta_1) \label{5}
\end{equation}

\noindent
where $ { \cal V}_{a,b}( \theta) $ denotes the two dimensional
space where the irrep (\ref{ 2 }) is defined.
Strictely speaking the "intertwiner matrix"
$ S^{k \; \ell}_{i \; j} \left(
\begin{array}{cc} a  & d  \\ b   & c  \end{array}
\right)$ will only exist for those sets of critical points
${a,b,c,d}$ such that the pairs $ (a,b), (b,c), (a,d) $
and $(d,c)$ are links of the LG-graph and satisfy the
parallelogram rule which guarantees an ellastic scattering:

\begin{eqnarray}
m_{a b} = m_{d c} &, & m_{b c} = m_{a d} \label{6}
\end{eqnarray}

After these definitions we can now write the
equation satisfied by the intertwiner operator (\ref{5}):

\begin{equation}
S\left( \begin{array}{cc} a  & d  \\ b   & c  \end{array}
\right)(\theta_{12})
( \pi_{a,b}(\theta_1) \otimes \pi_{b,c}(\theta_2) )
\Delta (g) =
( \pi_{a,d}(\theta_2) \otimes \pi_{d,c}(\theta_1) )
\Delta (g)
S\left( \begin{array}{cc} a  & d  \\ b   & c  \end{array}
\right)(\theta_{12})
\label{7}
\end{equation}

\noindent
where $g$ is any element of the N=2 algebra.
It is quite clear from the definition (\ref{5},\ref{7}) that the intertwiner
$S\left( \begin{array}{cc} a  & d  \\ b   & c  \end{array}
\right)(\theta)$ is the natural candidate for becoming
a scattering S-matrix between
Bogomolnyi solitons (see fig.1 ). In this case
equation (\ref{7}) is the result of imposing the N=2 SUSY to
the scattering process. Equation (\ref{7}) is the LG
analogue of the familiar intertwiner condition for Hopf algebras.
For these algebras the two irreps appearing on both sides
of the equation differ simply by a permutation.
However for LG models this will not happen in general,i.e. the irrep
$ \pi_{a,b}$ ( resp. $\pi_{b,c} )$ cannot be identified a priori with
the irrep $\pi_{d,c}$ ( resp. $\pi_{a,d}) $ eventhough the masses are the
same ( \ref{6}). In this manner the LG models suggests a generalization
of quantum groups where the LG-graph plays a crucial role. For
this reason we shall use the name of graph quantum groups to denote
the algebraic structure underlying these models.

The first equation defining a graph quantum group is therefore
the intertwiner condition (\ref{7}). As it is the case for
usual quantum groups this equation has to be supplemented with
what we shall call the graph-Yang-Baxter equation (see fig.2):

\begin{eqnarray}
\sum_f \sum_{p_1,p_2,p_3} &
S^{p_1 p_2}_{i_1 i_2} \left( \begin{array}{cc} e & f \\ a & b \end{array}
\right)(u)
S^{p_3 j_3}_{p_2 i_3} \left( \begin{array}{cc} f  & d \\ b & g \end{array}
\right)(u+v )
S^{j_1 j_2}_{p_1 p_3} \left( \begin{array}{cc} e & c \\ f & d \end{array}
\right)(v) = & \nonumber \\
\sum_f \sum_{p_1,p_2,p_3} &
S^{p_2 p_3}_{i_2 i_3} \left( \begin{array}{cc} a & f \\ b & g \end{array}
\right)(v)
S^{j_1 p_1}_{i_1 p_2} \left( \begin{array}{cc} e  & c \\ a & f \end{array}
\right)(u+v )
S^{j_2 j_3}_{p_1 p_3} \left( \begin{array}{cc} c & d \\ f & g \end{array}
\right)(u)  & \label{8}
\end{eqnarray}

In eq.(\ref{8}) there are two kinds of sums involved.
One is over the N=2 indices $p_1,p_2,p_3$ which is only
restricted by the conservation of the
fermion number , as in some vertex models, while
the sum over the LG label $f$ is restricted by the parallelogram
rule (\ref{6}) and recalls clearly the RSOS models
\footnote{ In eq.(\ref{8}) we are assuming that there
exist at most one kind of soliton , i.e. only
one link or none,
interpolating between two nodes of the graph}.Indeed eq.(\ref{8})
looks like the standard RSOS Yang-Baxter equation but with
Boltzmann weights replaced by four by four matrices whose indices are
contracted in the way prescribed by the vertex Yang-Baxter equation.
In this way the LG models produce a "fusion" of both the RSOS
and the vertex Yang-Baxter equations. The existence of solutions
to this "fused" Yang-Baxter equation associated to a LG-graph
is what guarantees the integrability of the corresponding
LG model.

In the cases studied so far
in the literature the solutions to  eq. (\ref{8}) factorize
into a vertex and a RSOS part each one satisfying his own Yang-Baxter
equation. The main result of this letter is to find solutions
which do not have this factorized form.Before doing that let
us review the previously known solutions to eq. (\ref{8}).

\subsection*{Solutions to the graph-Yang-Baxter equation}

i)$t_1$-perturbations of the A-models

The superpotential is given by :$W(x,t) = \frac{x^{k+2}}{k+2} - t x
$.The critical points are the $k+1$ roots of unit
$x_a = exp(2 \pi {\rm i} a/k+1) (t=1)$ with $a=1,
\dots, k+1$. The allowed parallelograms
$\left( \begin{array}{cc} a & d \\ b & c  \end{array} \right)$
are those satisfying $a-b = d-c$
and $a-d =b-c$ \cite{Fe} .Eq.(\ref{8})
becomes a standard vertex Yang-Baxter equation with
no summation over the LG indices. In this case it is even possible
to map the solitons irreps (\ref{ 2 }) into irreps of the
quantum affine algebra $U_q(\hat{Gl}(1,1))$ with
$q = e^{2 \pi {\rm i}/k +1}$ \cite{Fe} or equivalently to nilpotents irreps
of $U_q(\hat{Sl}(2))$ with $q^4 =1$ \cite{Ma} .The R matrices of these
quantum groups are automatically solutions of eq. (\ref{8}).
The physical S-matrix is finally obtained impossing bootstrap
which in this case turns out to coincide with the decomposition rules
of the quantum affine algebras and the factorization properties
of the R matrices.

ii) $t_k$-perturbation of $W(x) =\frac{x^{k+2} }{ k+2 }$

This is the well known Chebyshev potential of type A.
The LG-graph is the Coxeter diagram $A_{k+1}$ (see fig.
3b) and since
all the solitons have the same mass there is a sum
over the LG labels in eq. (\ref{8}).However the solution
to the later equation factorizes into the product \cite{Fe}:

\begin{equation}
S^{k \; \ell}_{i \; j} \left(
\begin{array}{cc} a  & d   \\ b   & c   \end{array}
\right)(\theta) =
S^{(RSOS)}\left(
\begin{array}{cc} a  & d   \\ b   & c   \end{array}
\right)(\theta) \otimes
S^{(N=2) k \; \ell}_{\;\;\;\;\;\;\;\;\;\;\ i \; j}
(\theta)
\label{10}
\end{equation}

\noindent
where $S^{(N=2)}$ is the intertwiner R matrix of the fundamental
representation of $U_q(\hat{Sl}(2))$  with $q^4 =1$ and
$S^{(RSOS)}$ are the ABF Boltzmann weights of the A models
at criticality \cite{ABF}. The later weights can also be writen in terms
of the $q-6j$ symbols of $U_q(Sl(2))$ with $q^{k+2} =-1$.
The origin of the factorization (\ref{10}) lies in the fact
that the ratios between different entries of the matrix
$S^{k \; \ell}_{i \; j} \left(
\begin{array}{cc} a  & d   \\ b   & c   \end{array}
\right) $,as predicted by the intertwiner eq.(\ref{7}),
are independent of the LG-labels. We shall next consider
a situation where this "separation of degrees of freedom"
does not occur so that the structure of the graph-Yang-Baxter
appears in fully glory.

\subsection*{$t_2$-perturbation of $W(x) =\frac{ x^{k+2}}{k+2}$}

The LG model is defined by the superpotential:$
W(x,t) =\frac{ x^{k+2}}{k+2} - t x^2$.The LG-graph of fundamental solitons has
the daisy shape of fig.3c
with a central point, denoted by a star ${*}$,
surrounded by k points lying on a circle
which we label by $a =1,\cdots,k$. For practical
purposes we shall restrict ourselves to odd values of $k$.
Equation (\ref{8}) for the graph of figure 3c becomes:

\begin{eqnarray}
R(\theta_1,\theta_2,\theta_3) \sum_{p_1,p_2,p_3}&
S^{p_1 p_2}_{i_1 i_2} \left( \begin{array}{cc} a & {*} \\ {*} & b \end{array}
\right)(\theta_{12})
S^{p_3 j_3}_{p_2 i_3} \left( \begin{array}{cc} {*}  & c \\ b & {*} \end{array}
\right)(\theta_{13})
S^{j_1 j_2}_{p_1 p_3} \left( \begin{array}{cc} a & * \\ {*} & c \end{array}
\right)(\theta_{23}) = & \nonumber \\
\sum_d \sum_{p_1,p_2,p_3} &
S^{p_2 p_3}_{i_2 i_3} \left( \begin{array}{cc} {*} & d \\ b & {*} \end{array}
\right)(\theta_{23})
S^{j_1 p_1}_{i_1 p_2} \left( \begin{array}{cc} a  & {*} \\ {*} & d \end{array}
\right)(\theta_{13})
S^{j_2 j_3}_{p_1 p_3} \left( \begin{array}{cc} {*} & c \\ d & {*} \end{array}
\right)(\theta_{12})  & \label{12}
\end{eqnarray}

We have slightly modified eq.(\ref{8}) with the introduction
of the factor $R(\theta_1,\theta_2,\theta_3)$. It will play
an important role in the discussion of the crossing symmetry
in the next section.
We are forced, as for the Chebyshev potential,
to sum over the LG labels, however the novelty here is that
the solution to (\ref{12}) cannot factorized into a RSOS
piece times a N=2 piece. Such a separation of degrees of freedom
is not possible in our case because , as follows from the intertwiner
condition (\ref{7}), the ratios between different entries
of the S-matrix does depend on the LG-labels.We collect our
results in table 1:

\begin{table}[h]
\begin{center}
\begin{tabular}{|c|c|c|}
\hline
$\begin{array}{cc} k & \ell \\ i & j \end{array}$ &
$S^{k \; \ell}_{i \; j} \left(
\begin{array}{cc} a  & {*}   \\ {*}   & b   \end{array}
\right) /
S^{0 \; 0 }_{0 \; 0} \left(
\begin{array}{cc} a  & {*}   \\ {*}   & b   \end{array}
\right)$ &
$S^{k \; \ell}_{i \; j} \left(
\begin{array}{cc} {*}  & {b}   \\ {a}   & {*}   \end{array}
\right) /
S^{0 \; 0 }_{0 \; 0} \left(
\begin{array}{cc}  {*} & b  \\ a &  {*}     \end{array}
\right)$ \\ \hline
$\begin{array}{cc} 1 & 1   \\ 1 & 1 \end{array}$ &
$\frac{cosh( \frac{\theta}{2}+\frac{2 \pi {\rm i} n}{k})
}{cosh( \frac{\theta}{2}-\frac{2 \pi {\rm i} n}{k}) }$  &
$exp(-\frac{4 \pi {\rm i} n}{k})$ \\ \hline
$\begin{array}{cc} 1 & 0    \\ 0 & 1 \end{array}$ &
$-\frac{{\rm i} exp(2 \pi {\rm i} n/k) sinh(\frac{\theta}{2})
}{cosh( \frac{\theta}{2}-\frac{2 \pi {\rm i} n}{k}) } $ &
$-\frac{{\rm i} exp(-2 \pi {\rm i} n/k)
sinh(\frac{\theta}{2} -\frac{2 \pi {\rm i} n}{k})
}{cosh \frac{\theta}{2} }$ \\ \hline
$\begin{array}{cc} 0 & 1    \\ 1 & 0 \end{array}$ &
$-\frac{  {\rm i} exp(-2 \pi {\rm i} n/k) sinh(\frac{\theta}{2})
}{cosh( \frac{\theta}{2}-\frac{2 \pi {\rm i} n}{k}) } $ &
$-\frac{{\rm i} exp(-2 \pi {\rm i} n/k)
sinh(\frac{\theta}{2} +\frac{2 \pi {\rm i} n}{k})
}{cosh \frac{\theta}{2} }$ \\ \hline
$\begin{array}{cc} 1 & 0    \\ 1 & 0 \end{array}$ &
$\frac{cos( \frac{2 \pi n}{k})
}{cosh( \frac{\theta}{2}-\frac{2 \pi {\rm i} n}{k}) }$  &
$\frac{exp(-2 \pi {\rm i} n/k) cos(\frac{2 \pi n}{k} )
}{cosh \frac{\theta}{2} }$  \\ \hline
$\begin{array}{cc} 0 & 1    \\ 0 & 1 \end{array}$ &
$\frac{cos( \frac{2 \pi n}{k})
}{cosh( \frac{\theta}{2}-\frac{2 \pi {\rm i} n}{k}) } $ &
$\frac{exp(-2 \pi {\rm i} n/k) cos(\frac{2 \pi n}{k} )
}{cosh \frac{\theta}{2} }$ \\ \hline
\end{tabular}
\caption{Relations imposed by the N=2 intertwiner
condition}
\end{center}
\label{1}
\end{table}

\noindent
where $n=a-b$. Notice that all the expressions depend
explicitely on the LG labels modulo $k$.
Each of these sets of matrices satisfy the free fermion
condition \cite{FF}:
$S^{0 0}_{0 0} S^{1 1}_{1 1}=S^{1 0}_{0 1} S^{0 1}_{1 0} +
S^{1 0}_{1 0} S^{0 1}_{0 1}$
,which is of course a
consequence of the N=2 SUSY.

The $t_2$ perturbation is the simplest LG model
which imposes the search of non factorizable solutions
,in the sense explained above, of the graph-Yang-Baxter
equation.
We now move on to present the solution of the graph-Yang-Baxter
equation (\ref{12}).The key for solving this equation
comes from its similarity with the star-triangular equation of the
Chiral-Potts model \cite{CP}. This analogy becomes more transparent
in the IRF version of the $Z_k$ invariant Chiral-Potts model,
since the neutral element under the $Z_k$ symmetry can be identified
with the star point ${*}$ in the LG graph. Moreover the two kinds
of S-matrices namely
$S \left(
\begin{array}{cc} a  & {*}   \\ {*}   & b   \end{array}
\right)$ and
$S \left(
\begin{array}{cc} {*}  & {b}   \\ {a}   & {*}   \end{array}
\right)$ , which described respectively
soliton-antisoliton and antisoliton-soliton scatterings,
correspond to horizontal and vertical Boltzmann weights of the
Chiral-Potts model. A particular case of eq.(\ref{12}) is when all
the N=2 indices $i,j,etc.$ are either 0 or 1. Then eq.(\ref{12})
becomes the star-triangular equation of Chiral-Potts
for the entries $S^{0 0}_{0 0}$ or $S^{1 1}_{1 1}$.
There is a great variety of solutions to this equation,
however our search is limited by the fact that the ratio of
the two solutions $S^{0 0}_{0 0}$ and  $S^{1 1}_{1 1}$
must satisfy the constraints of table 1.
The unique solution we find to this problem
is given by:

\begin{eqnarray}
S^{0 \; 0 }_{0 \; 0} \left(
\begin{array}{cc} a  & {*}   \\ {*}   & b   \end{array}
\right)(\theta) &=&
S(\theta) \prod^{a-b-1}_{r=0}
\frac{ cosh( \frac{\theta}{2} + \frac{2 \pi {\rm i} r}{k}) }{
cosh ( \frac{\theta}{2} - \frac{2 \pi {\rm i} r}{k})} \label{14} \\
S^{0 \; 0 }_{0 \; 0} \left(
\begin{array}{cc} {*}  & {b}   \\ {a}   & {*}   \end{array}
\right)(\theta) &=&
\bar{S}(\theta)
e^{2 \pi {\rm i} (a-b)/k} \;\prod^{a-b-1}_{r=0}
\frac{ sinh( \frac{\theta}{2} - \frac{2 \pi {\rm i} r}{k}) }{
sinh ( \frac{\theta}{2} + \frac{2 \pi {\rm i} (r+1)}{k})} \nonumber
\end{eqnarray}

\noindent
where $S(\theta)$ and $\bar{S}(\theta)$ denote:

\begin{eqnarray}
S(\theta)=S^{0 \; 0 }_{0 \; 0} \left(
\begin{array}{cc} a  & {*}   \\ {*}   & a   \end{array}
\right)(\theta),& &
\bar{S}(\theta)=
S^{0 \; 0 }_{0 \; 0} \left(
\begin{array}{cc} {*}  & {a}   \\ {a}   & {*}   \end{array}
\right)(\theta)
\label{15}
\end{eqnarray}

\noindent
Eqs.(\ref{14}) are $Z_k$ invariant, i.e. depend on
a-b modulo k, and they satisfy the star-triangular equation
with the R factor being given by:

\begin{eqnarray}
R(\theta_1,\theta_2, \theta_3)& = & \frac{ f(\theta_{12}) f(\theta_{23})}{
f(\theta_{13})}
\nonumber \\
f(\theta)&  = & \frac{ \bar{S}(\theta)}{S(\theta)}
\sum^k_{m=1} e^{ 2 \pi {\rm i} m/k}
\;\; \prod^{m-1}_{r=0}
\frac{ sinh( \frac{\theta}{2} - \frac{2 \pi {\rm i} r}{k}) }{
sinh ( \frac{\theta}{2} + \frac{2 \pi {\rm i} (r+1)}{k})} \label{17}
\end{eqnarray}

The solution (\ref{14}) and the corresponding one for
$S^{1 1}_{1 1}$ coincide in fact with the Boltzmann
weights of
a $Z_k$ invariant Chiral-Potts model
characterized by a parameter $\omega=e^{4 \pi {\rm i}/k}$,
"moduli" $k'=1$
and "chiral angles":

\begin{equation}
\phi = \frac{ \pi }{ 2 } (k \mp 2),\;\;
\bar{\phi} = \frac{ \pi }{ 2 }  (k \pm 2)\;\;
{\rm for}\;\;S^{0 0}_{0 0} \;\; (S^{1 1}_{1 1})   \label{18}
\end{equation}

For k odd the parameter $\omega$ is a
$k^{th}$ primitive root
of unit while for k even
the degree of the root is $\frac{k}{2}$.
Hence we expect certain
differences for the
$k$ even case as compared to the
$k$ odd case.Eq.(\ref{18}) suggests some kind of relation
of these models with the
superintegrable models of reference \cite{Va} .

Once we know the values of $S^{0 0}_{0 0}$ the rest of the
entries of the matrix S follow from
table 1.
This huge S-matrix does satisfy the graph-Yang-Baxter
equation (\ref{12}), a fact which by no means follows
a priori from the existence of the partial solutions
(\ref{14}). This "miracle" is certainly due to the
integrability of the LG model under study.
To get a feeling of the intricacy of the graph-Yang-Baxter
equation it is quite useful to consider the
braid limit ($\theta \rightarrow \infty$) of the
solution just found. The braid matrix
so obtained is given in table 2 and satisfies the
graph-Yang-Baxter equation without spectral parameter
(we have choosen a parametrization where the parameter
R is given by the gaussian sum $R(braid)= \sum^k_{n=1}
q^{-n^2 +n}$  ).

\begin{table}[h]
\begin{center}
\begin{tabular}{|c|c|c|}
\hline
$\begin{array}{cc} k & \ell \\ i & j \end{array}$ &
$R^{k \; \ell}_{i \; j} \left(
\begin{array}{cc} a  & {*}   \\ {*}   & b   \end{array}
\right) $ &
$R^{k \; \ell}_{i \; j} \left(
\begin{array}{cc} {*}  & {b}   \\ {a}   & {*}   \end{array}
\right) $ \\ \hline
$\begin{array}{cc} 0 & 0   \\ 0 & 0 \end{array}$ &
$q^{n^2 - n}$ & $q^{-n^2 +n}$ \\ \hline
$\begin{array}{cc} 1 & 1   \\ 1 & 1 \end{array}$ &
$q^{n^2 +n}$ & $q^{-n^2 -n}$ \\ \hline
$\begin{array}{cc} 1 & 0    \\ 0 & 1 \end{array}$ &
$- {\rm i} q^{n^2 +n}$ & $ - {\rm i} q^{-n^2 -n}$ \\ \hline
$\begin{array}{cc} 0 & 1    \\ 1 & 0 \end{array}$ &
$- {\rm i} q^{n^2 - n}$ & $- {\rm i} q^{-n^2 +n}$ \\ \hline
$\begin{array}{cc} 1 & 0    \\ 1 & 0 \end{array}$ &
$q^{n^2} (q^n + q^{-n})$ &$q^{-n^2} (q^n + q^{-n})$   \\ \hline
$\begin{array}{cc} 0 & 1    \\ 0 & 1 \end{array}$ &
$0$ & $0$ \\ \hline
\end{tabular}
\caption{A graph braid R-matrix }
\end{center}
\label{2}
\end{table}

\noindent
where $n =a -b$ and $q=\omega^{1/2}=e^{2 \pi {\rm i}/k}$.

This concludes our discussion of the integrability of the
$t_2-A LG$ model.
In the next section we shall discuss the
interpretation of the solution found previously as scattering
matrices of the solitons of this model.

\section*{Scattering Theory}

A good scattering theory requires crossing symmetry which in
our case amounts to the relation:

\begin{equation}
S^{k \; \ell}_{i \; j} \left(
\begin{array}{cc} {*}  & {b}   \\ {a}   & {*}   \end{array}
\right)(\theta) =
S^{\bar{\imath} \; k}_{j \; \bar{\ell}} \left(
\begin{array}{cc} a  & {*}   \\ {*}   & b   \end{array}
\right)({\rm i} \pi - \theta)
\label{19}
\end{equation}

\noindent
where we have assumed that the antiparticle of the soliton
(a,${*}$) with SUSY label $i$ = 0,1 is the antisoliton
(${*}$,a) with
SUSY label $\bar{\imath}$ =1,0.
Using table 1 and eq.(\ref{14}) we obtain that crossing
symmetry holds provided:

\begin{equation}
S({\rm i} \pi - \theta) = - {\rm i} tanh(\frac{\theta}{2}) \;\;\;
\bar{S} (\theta)
\label{20}
\end{equation}

This equation together with (\ref{17}) implies the following
interesting relation for the function $f(\theta)$:

\begin{equation}
f(\theta) \;\; f({\rm i} \pi - \theta) = k
\label{21}
\end{equation}

\noindent
whose main consequence is that the factor
$R(\theta_1,\theta_2,\theta_3)$
cannot be set equal to one , as one could naively expect.
A reasonable choice is $R = f(\theta)= \sqrt{k}$ so that
$R$ can be viewed as a counting factor which compensates
the asymmetry present in the graph-Yang-Baxter equation (\ref{12}).
This phenomena presumably happens also for other LG graphs
with a daisy shape.

A second strange fact appears in connection with the unitarity
properties of the S-matrices. Indeed unitarity implies
in our case:

\begin{eqnarray}
S\left(
\begin{array}{cc} a  & {*}   \\ {*}   & b   \end{array}
\right)(\theta)
S\left(
\begin{array}{cc} a  & {*}   \\ {*}   & b   \end{array}
\right)(-\theta)= &  {\bf 1}& \label{22} \\
\sum_b
S\left(
\begin{array}{cc} {*}  & {a}   \\ {b}   & {*}   \end{array}
\right)(\theta)
S\left(
\begin{array}{cc} {*}  & {b}   \\ {c}   & {*}   \end{array}
\right)(-\theta) = & \delta_{a,c} \nonumber
\end{eqnarray}

\noindent
However it is not difficult to see that these two equations
are incompatible with crossing, which indicates that
the scattering theory is not complete.The solution of
this problem requires a careful analysis of the
spectrum of bound states and the bootstrap equations
that we postpone to a later paper. In any case one expects
the bootstrap to work, just as Yang-Baxter does, in a highly
non trivial way, due again to the interplay between various
degrees of freedom.

In this letter we show that the
understanding of integrable N=2 LG models
requires new mathematical structures
which are in a certain sense a
synthesis of well established objects.In this
direction we introduced the notion of
graph quantum groups which leads naturally
to the graph-Yang-Baxter equation.
These ideas are illustrated with the
study of the $t_2$ perturbation of the
A-models where the unification of
structures ,in the sense explained
above ,occurs for the first time.

\vspace{1cm}
{\bf Acknowledgments}

We would like to thanks M.Ruiz--Altaba for discussions.
The research of  C.Gomez
is supported in part
by the Swiss National Science
Foundation.

\newpage

\section*{Figure Captions}

\noindent
Fig.1.- Graphic representation of the intertwiner S-matrix.

\noindent
Fig.2.- Graph-Yang-Baxter equation.

\noindent
Fig.3.- Graphs associated to the following
perturbations of the type A Landau-Ginzburg models:
$t_1 \leftrightarrow a, t_k \leftrightarrow b$ and
$t_2 \leftrightarrow c$.

\end{document}